\documentclass[conference]{IEEEtran}
\IEEEoverridecommandlockouts

\usepackage{cite}
\usepackage{amsmath,amssymb,amsfonts,amsthm}
\usepackage{mathabx}
\usepackage{stmaryrd}
\usepackage{multirow}
\usepackage{float}
\usepackage{hyperref}
\usepackage{graphicx}
\usepackage{textcomp}
\usepackage{algorithm, algpseudocode}
\usepackage{xcolor,tikz, circuitikz}
\def\BibTeX{{\rm B\kern-.05em{\sc i\kern-.025em b}\kern-.08em
    T\kern-.1667em\lower.7ex\hbox{E}\kern-.125emX}}

\newcommand{\ket}[1]{|{#1}\rangle}
\newcommand{\bra}[1]{\langle{#1}|}

\newtheorem{theorem}{Theorem}

\newtheorem{definition}{Definition}

\begin{document}

\title{Quantum Approximate Optimization for Decoding of Low-Density Parity-Check Codes}

\author{\IEEEauthorblockN{1\textsuperscript{st} Krishnakanta Barik}
\IEEEauthorblockA{\textit{Indian Statistical Institute}\\
Kolkata 700108, India \\
krishnakanta\_r@isical.ac.in}
\and
\IEEEauthorblockN{2\textsuperscript{nd} Goutam Paul}
\IEEEauthorblockA{\textit{Indian Statistical Institute}\\
Kolkata 700108, India \\
goutam.paul@isical.ac.in}
}

\maketitle

\begin{abstract}
Decoding Low-Density Parity-Check (LDPC) codes is a fundamental problem in coding theory, and  Belief Propagation (BP) is one of the most popular methods for LDPC code decoding.
However, BP may encounter convergence issues and suboptimal performance, especially for short-length codes and in high-noise channels. The Quantum Approximate Optimization Algorithm (QAOA) is a type of Variational Quantum Algorithm (VQA) designed to solve combinatorial optimization problems by minimizing a problem-specific cost function.
In this paper, we present a QAOA-based decoding framework for LDPC codes by formulating a decoding cost function that incorporates both parity-check constraints and soft channel reliability information. 
The resulting optimization problem is solved using QAOA to search for low-energy configurations corresponding to valid codewords. 
We test the proposed method through extensive numerical experiments and compare its performance with BP decoding.
The experimental results demonstrate that the QAOA-based decoder achieves a higher probability of correctly recovering the transmitted codeword than BP across multiple experimental settings.
\end{abstract}

\section{INTRODUCTION}
A code~\cite{richardson2008modern} is a structured set of bit strings, usually called codewords, constructed according to predefined rules to enable reliable transmission over a communication channel. In error-correcting codes, redundancy is deliberately added to the transmitted data. This additional structure allows the receiver to recover the original information even when the channel introduces noise~\cite{1057683}.
One widely used example of such codes is the family of LDPC codes~\cite{1057683, richardson2008modern}.
Because of their strong error-correction performance and the availability of efficient decoding algorithms, LDPC codes are used in modern communication standards such as Wi-Fi and 5G~\cite{richardson2008modern}.
A decoder~\cite{richardson2008modern} is an algorithm whose task is to estimate the transmitted codeword from the noisy channel output. 
For LDPC codes, the most commonly used decoding method is Belief Propagation (BP)~\cite{richardson2008modern}, which is an iterative message-passing algorithm that updates the reliability values associated with each bit. Although BP is exact for graphs without cycles, practical LDPC codes typically contain cycles. As a result, BP may fail to converge or may produce incorrect decoding results, even at moderate noise levels~\cite{richardson2008modern, 748992}.

Quantum computing~\cite{nielsen2010quantum} exploits quantum-mechanical principles like entanglement and superposition to perform computations that are difficult or infeasible for classical computers in certain problem settings. In recent years, considerable attention has been given to quantum algorithms suitable for near-term, noisy quantum devices, commonly referred to as Noisy Intermediate-Scale Quantum (NISQ) systems \cite{preskill2018quantum}. This interest is strong in the context of combinatorial optimization problems~\cite{korte2008combinatorial}.
One prominent example is the Quantum Approximate Optimization Algorithm (QAOA)~\cite{farhi2014quantumapproximateoptimizationalgorithm, a12020034}. QAOA is a VQA~\cite{cerezo2021variational, mitarai2018quantum, wecker2015progress} that aims to find approximate solutions to optimization problems by alternately applying problem-specific quantum operators and mixing quantum operators.
Because of its hybrid quantum–classical structure and its natural suitability for binary optimization problems, the QAOA has been applied to a broad range of domains. These include graph optimization~\cite{PhysRevApplied.19.024027}, the tail assignment problem~\cite{PhysRevApplied.14.034009}, and large sparse hypergraphs and spin glass models~\cite{9996946}.

In this work, we examine the potential of quantum computing in improving the classical decoding process for linear block codes. 
We define a decoding cost function that takes into account both the parity-check matrix of the code and the reliability information of the noisy communication channel. This cost function provides a natural formulation of the decoding problem as an optimization problem over the codeword space.
Based on this cost function, we develop a QAOA decoding scheme that seeks to find low-energy states representing valid codewords.
Numerous experiments are performed to test the performance of the proposed scheme. In nine different experimental settings, each consisting of around 100 independent decoding attempts, the QAOA-based decoder outperforms the classical BP decoder in terms of the empirical decoding success probability, especially in cases where the BP decoder experiences convergence problems.

We organized this paper as follows. Section~II introduces the background material needed for this work.
In particular, we review key concepts from classical coding theory, including linear block codes, parity-check matrices, and Belief Propagation (BP) decoding. We also provide the necessary quantum computing background, with an emphasis on the Quantum Approximate Optimization Algorithm (QAOA).
In Section~III, we present the proposed QAOA-based decoding framework. 
This section describes the construction of the decoding cost function, explains how it incorporates parity-check constraints and channel reliability information, and outlines how QAOA is applied to the decoding task.
Section~IV describes the experimental numerical results.
Here, the performance of the proposed QAOA-based decoder is evaluated and compared with classical BP decoding across different noise conditions.
In Section~V, we discuss the benefits of using QAOA as a decoding approach, followed by the conclusion and future directions in Section~VI.


\section{Preliminaries}

In this section, we present the necessary classical and quantum preliminaries needed for the proposed decoding framework. 
Specifically, we review the fundamental concepts from coding theory and quantum computation that form the basis of the decoder model.

\subsection{Classical Preliminaries}

In classical coding theory, a code is a structured collection of sequences used to represent information for reliable transmission or storage. 
The formal definition of a code is given below.

\begin{definition}[Code~\cite{richardson2008modern}]
A code \( C \) of length \( n \) and cardinality \( M \) over a field \( \mathbb{F} \) is a collection of \( M \) elements from \( \mathbb{F}^n \), i.e.,
\begin{align*}
C(n,M) := \{ \mathbf{x}^{(1)}, \ldots, \mathbf{x}^{(M)} \},
\mathbf{x}^{(m)} \in \mathbb{F}^n,\; 1 \le m \le M.    
\end{align*}
The elements of \( C \) are called codewords, and the parameter \( n \) is referred to as the block-length.
\end{definition}

Before formally defining Low-Density Parity-Check (LDPC) codes, we first introduce the notions of linear codes, parity-check matrices, and Tanner graphs.

\begin{definition}[Linear Code~\cite{richardson2008modern}]
A code \( C \subseteq \mathbb{F}^n \) over a field \( \mathbb{F} \) is called linear if it is closed under vector addition and scalar multiplication, that is,
\[
\alpha \mathbf{x} + \beta \mathbf{x}' \in C,
\quad \forall\, \mathbf{x}, \mathbf{x}' \in C,\; \forall\, \alpha, \beta \in \mathbb{F}.
\]
\end{definition}

A linear code \( C \) of length \( n \) over a field \( \mathbb{F} \) forms a vector subspace of \( \mathbb{F}^n \) and therefore has a well-defined dimension \( k \), with \( 0 \le k \le n \).
A generator matrix \( G \) is a \( k \times n \) matrix whose rows constitute a linearly independent basis for the code \( C \).

\begin{definition}[Parity-Check Matrix~\cite{richardson2008modern}]
Let \( C \subseteq \mathbb{F}^n \) be a linear code with generator matrix \( G \).
The dual code of \( C \) is defined as
\[
C^\perp := \{ \mathbf{v} \in \mathbb{F}^n \mid \mathbf{x}\mathbf{v}^T = 0,\ \forall\, \mathbf{x} \in C \}.
\]
A matrix \( H \) whose rows form a basis of the dual code \( C^\perp \) is called a
parity-check matrix of the code \( C \).
\end{definition}
Thus, if \( H \) is a parity-check matrix of a linear code \( C \), then the code can be expressed equivalently as
\[
C := \{ \mathbf{x} \in \mathbb{F}^n \mid H \mathbf{x}^T = \mathbf{0}^T\}.
\]

\begin{definition}[Tanner Graph~\cite{richardson2008modern}]
Let \( C \) be a binary linear code and let \( H \) be a parity-check matrix of \( C \).
By convention, the rows of \( H \) are not required to be linearly independent.
Assume that \( H \) has dimensions \( m \times n \).
The Tanner graph associated with \( H \) is a bipartite graph consisting of two types of nodes:
\begin{itemize}
    \item \( n \) variable nodes, corresponding to the components of a codeword, and
    \item \( m \) check nodes, corresponding to the parity-check constraints (rows of \( H \)).
\end{itemize}
A check node \( j \) is connected to a variable node \( i \) if and only if \( H_{ji} = 1 \), that is, if variable \( i \) participates in the \( j \)-th parity-check constraint.
\end{definition}

Since a given linear code may admit multiple parity-check matrices, there can be multiple Tanner graphs associated with the same code \( C \).

\begin{definition}[Low-Density Parity-Check Codes~\cite{richardson2008modern}]
LDPC codes are linear codes that admit at least one sparse Tanner graph.
\end{definition}
In the remainder of this section, we outline the decoding problem for linear codes and briefly review the BP algorithm, which is the most commonly used decoding method for LDPC codes.
A decoder can be viewed as an algorithm tasked with inferring the transmitted codeword from a noisy channel observation while satisfying the parity-check constraints imposed by the code.
Given a received vector, the decoder aims to identify a codeword that is consistent both with the channel output and with the underlying algebraic structure of the code.
A key quantity in decoding theory, and in the BP algorithm, is the Log-Likelihood Ratio (LLR)~\cite{richardson2008modern}. Channel reliability information is represented using this ratio. For the \(i\)-th bit, the LLR is defined as
\[
\mathrm{LLR}_i:
=
\log \frac{p(y_i \mid c_i = 0)}{p(y_i \mid c_i = 1)},
\]
where \(y_i\) denotes the received channel output corresponding to the transmitted bit \(c_i\). A positive LLR indicates that the transmitted bit is more likely to be \(0\), whereas a negative LLR favors the bit value \(1\).

BP is an iterative message-passing decoding algorithm defined on the Tanner graph of a code.
In BP, variable and check nodes communicate by passing probabilistic messages that reflect beliefs about the transmitted bits.
These messages are updated locally using the parity-check constraints and the channel observations.
BP aims to approximate the marginal posterior probabilities of the codeword bits and to estimate the most likely transmitted codeword.
The message passed from a variable node \( v \) to a check node \( c \) is defined as
\[
m_{v \rightarrow c}
:=
\mathrm{LLR}_v
+
\sum_{c' \in N(v)\setminus c}
m_{c' \rightarrow v},
\]
\( \mathrm{LLR}_v \) denotes the channel log-likelihood ratio associated with variable node \( v \), and \( N(v) \) represents the set of check nodes connected to \( v \).
The message sent from a check node \( c \) to a variable node \( v \) is computed as
\[
m_{c \rightarrow v}
:=
2 \tanh^{-1}
\left(
\prod_{v' \in N(c)\setminus v}
\tanh\left( \frac{m_{v' \rightarrow c}}{2} \right)
\right),
\]
where \( N(c) \) denotes the set of variable nodes connected to check node \( c \).
After a fixed number of iterations or upon convergence, the final log-likelihood ratio for bit \( i \) is given by
\[
\mathrm{LLR}_i^{\mathrm{final}}
=
\mathrm{LLR}_i
+
\sum_{c \in N(i)}
m_{c \rightarrow i}.
\]
The decoded bit \( \hat{x}_i \) is obtained using the hard-decision rule
\[
\hat{x}_i =
\begin{cases}
0, & \text{if } \mathrm{LLR}_i^{\mathrm{final}} > 0, \\
1, & \text{if } \mathrm{LLR}_i^{\mathrm{final}} < 0.
\end{cases}
\]
BP may exhibit poor decoding performance when the Tanner graph of the code contains short cycles, which introduce strong dependencies among messages and violate the independence assumptions underlying the algorithm. 
In such situations, messages can repeatedly reinforce incorrect beliefs, leading to unreliable updates and convergence to suboptimal solutions. Another prominent failure mode arises from the presence of trapping sets or stopping sets, which are small subsets of variable nodes that satisfy a large fraction of the parity-check constraints while remaining globally inconsistent with the transmitted codeword. Once BP enters these configurations, it often becomes trapped and fails to converge to the correct solution. 
In addition, BP can perform poorly when the channel provides strong but incorrect reliability information, causing the log-likelihood ratios to dominate the parity-check constraints and bias the decoder toward an incorrect codeword.

Several alternative decoding algorithms have been proposed for linear and LDPC codes. 
Maximum Likelihood (ML)~\cite{wolf1978efficient} decoding provides optimal decoding performance by selecting the codeword that maximizes the likelihood given the channel observation; however, its computational complexity grows exponentially with the code length, rendering it impractical for moderate and large block sizes. 
Simpler iterative decoders, such as the Min-Sum algorithm, reduce computational complexity by approximating the belief propagation updates, but typically suffer from reduced decoding accuracy. 
Bit-Flipping~\cite{wadayama2010gradient} decoders offer very low complexity by iteratively correcting bits that violate the most parity-check constraints; however, their performance is generally poor, particularly in the presence of noise and short cycles in the Tanner graph.

\subsection{Quantum Preliminaries}

Quantum Machine Learning (QML)~\cite{biamonte2017quantum} integrates ideas from quantum computing and machine learning to explore how quantum resources can improve learning and inference.
Within QML, VQAs~\cite{cerezo2021variational, mitarai2018quantum, wecker2015progress} are a class of hybrid quantum–classical methods in which a parameterized quantum circuit is optimized using a classical optimizer to minimize a problem-dependent cost function.
VQAs are particularly well-suited to near-term quantum hardware~\cite{cerezo2021variational}, as they employ parameterized quantum circuits while delegating the optimization task to classical computation, where errors arising from quantum circuits can be mitigated (absorbed) through iterative classical optimization steps.

The Quantum Approximate Optimization Algorithm (QAOA)~\cite{zhou2020quantum, farhi2014quantumapproximateoptimizationalgorithm, a12020034} is a well-known example of a variational quantum algorithm developed to approximately solve combinatorial optimization problems.
In QAOA, an optimization problem defined over binary variables is encoded into a problem-dependent cost Hamiltonian \( H_P \), whose ground state corresponds to the optimal solution. 
Starting from an initial quantum state \( \ket{s} \), QAOA evolves the system by alternately applying the cost Hamiltonian and a mixing Hamiltonian \( H_M \), which induces transitions between computational basis states.
For a circuit depth \( l \), the QAOA ansatz state is given by
\[
\ket{\boldsymbol{\gamma}, \boldsymbol{\beta}}
=
\prod_{j=1}^{l}
e^{-i \beta_j H_M}
e^{-i \gamma_j H_P}
\ket{s},
\]
where \( \boldsymbol{\gamma} = (\gamma_1, \ldots, \gamma_l) \) and \( \boldsymbol{\beta} = (\beta_1, \ldots, \beta_l) \) are variational parameters. 
The objective of QAOA is to minimize the expected weight of the cost Hamiltonian corresponding to this parameterized circuit, leading to the cost function
\[
P(\boldsymbol{\gamma}, \boldsymbol{\beta}):
=
\bra{\boldsymbol{\gamma}, \boldsymbol{\beta}} H_P \ket{\boldsymbol{\gamma}, \boldsymbol{\beta}}.
\]
The optimization of QAOA is carried out through a hybrid quantum-classical loop, in which a classical optimizer updates the parameters \( (\boldsymbol{\gamma}, \boldsymbol{\beta}) \) based on measurement outcomes.
At each iteration, the quantum device prepares the QAOA state and estimates the cost function through repeated measurements, while the classical optimizer seeks parameter values that reduce the expected energy. 
After convergence, measuring the optimized QAOA state in the computational basis yields candidate bit strings, with high probability assigned to near-optimal solutions of the original optimization problem.

\section{QAOA for LDPC Decoding}

As discussed in the previous section, the design of a QAOA for a given problem instance consists of four essential components: 
(i) a problem-dependent cost Hamiltonian \( H_P \) corresponding to an objective function over binary variables, 
(ii) a mixing Hamiltonian \( H_M \) that enables transitions between feasible configurations, 
(iii) an initial quantum state \( \ket{s} \), and 
(iv) a classical optimization procedure for updating the variational parameters.
In the QAOA-based LDPC decoder, we adopt the standard choice of the mixing Hamiltonian and initial state
\[
H_M := \sum_{j=1}^{n} X_j, \qquad \ket{s} := \ket{+}^{\otimes n}.
\]
where $X_j$ is the Pauli-$X$ operator acting on the $j$-th qubit.
To optimize the variational parameters, we employ the parameter-shift rule~\cite{bergholm2018pennylane,schuld2019evaluating}, which allows the gradients of the quantum circuit with respect to its tunable parameters to be evaluated analytically and facilitates efficient classical optimization.
The most critical component of the proposed framework is the construction of the cost Hamiltonian \( H_P \), which encodes both the parity-check constraints of the LDPC code and the channel reliability information.
The design of this Hamiltonian is described in detail in the remainder of this section.

We consider a binary vector
$
\mathbf{x} = (x_1, x_2, \ldots, x_n),
$
and map the binary variables to Ising~\cite{lucas2014ising} spin variables according to
$
Z_i = (-1)^{x_i}, Z_i \in \{+1,-1\}.
$
This representation enables a natural formulation of parity-check constraints and facilitates the construction of an Ising-type Hamiltonian whose ground state corresponds to a valid LDPC codeword with maximum likelihood.
Let \( \mathcal{C} \) denote the set of parity-check constraints of the linear code \( C \), by the parity-check matrix \( H \). For a given parity check \( c \in \mathcal{C} \), involving the set of variable nodes \( i \in c \), the corresponding parity constraint can be expressed as
$
\sum_{i \in c} x_i = 0 \pmod{2}.
$

Using the Ising-spin representation \( Z_i := (-1)^{x_i} \), this condition is equivalently expressed as 
\[
\prod_{i \in c} Z_i = 1.
\]
Based on this observation, we introduce a parity-check penalty term that assigns zero energy to configurations satisfying the parity constraint and a positive energy penalty otherwise. The resulting parity-check Hamiltonian is defined as
\[
H_{\text{parity}}(Z):
=
\sum_{c \in \mathcal{C}} \frac{1}{2}
\left( 1 - \prod_{i \in c} Z_i \right).
\]
This Hamiltonian evaluates to zero if and only if all parity checks are satisfied, and assigns a strictly positive energy to any configuration that violates at least one parity constraint.
Since the parity-check matrix of an LDPC code is sparse, the corresponding parity-check Hamiltonian \( H_{\mathrm{parity}}(Z) \) can be implemented efficiently on quantum devices~\cite{harrow2009quantum}.
The influence of channel noise is incorporated through a channel-dependent penalty term constructed from the Log-Likelihood Ratios (LLRs).
Let \( h_i \) denote the channel LLR associated with the \( i \)-th transmitted bit.
Using the Ising-spin representation \( Z_i \in \{+1,-1\} \), we define the channel penalty Hamiltonian as
\[
H_{\text{channel}}(Z):
=
\sum_{i=1}^{n}
\frac{|h_i|}{2}
\left( 1 - \operatorname{sign}(h_i)\, Z_i \right).
\]

This term encodes the channel preference for each bit based on the received observation.
Specifically, a positive LLR \( h_i > 0 \) indicates that the transmitted bit \( x_i = 0 \) is more likely, corresponding to the spin value \( Z_i = +1 \),
whereas a negative LLR \( h_i < 0 \) favors the bit value \( x_i = 1 \), corresponding to \( Z_i = -1 \).
The magnitude \( |h_i| \) quantifies the confidence of the channel observation and determines the strength of the associated energy penalty.

From an energy-minimization perspective, the contribution of the \(i\)-th term of the channel Hamiltonian is minimized if and only if 
\[
Z_i = \operatorname{sign}(h_i).
\]
Specifically, if \( h_i > 0 \), then \( \operatorname{sign}(h_i) = +1 \) and the energy contribution 
\[
\frac{|h_i|}{2}(1 - Z_i)
\]
is minimized when \( Z_i = +1 \). Similarly, if \( h_i < 0 \), then \( \operatorname{sign}(h_i) = -1 \) and the contribution 
\[
\frac{|h_i|}{2}(1 + Z_i)
\]
is minimized when \( Z_i = -1 \).
Consequently, the channel Hamiltonian assigns lower energy to bit configurations if and only if they are consistent with the channel observations, thereby guiding the optimization toward the most likely transmitted codeword.

The total cost Hamiltonian \( H(Z) \) for the QAOA-based LDPC decoder is constructed by combining the parity-check penalty and the channel-dependent penalty terms.
Specifically, the decoding Hamiltonian is defined by
\begin{align*}
H(Z):= & H_{\mathrm{parity}}(Z) + H_{\mathrm{channel}}(Z) \\
=      &\sum_{c \in \mathcal{C}} \frac{1}{2} \left(1 - \prod_{i \in c} Z_i \right) + \sum_{i=1}^{n}\frac{|h_i|}{2}
\left(1 - \operatorname{sign}(h_i)\, Z_i \right).
\end{align*}

The key insight underlying this formulation is that valid codewords correspond to low-energy configurations of the decoding Hamiltonian. By minimizing the expectation value of \( H(Z) \) using QAOA, the decoding task is naturally cast as an optimization problem, in which measured low-energy bit strings correspond to candidate decoded codewords. This observation leads to the following result.

\begin{theorem}
Any minimizer of \( H(Z) \) corresponds to a codeword that satisfies all parity-check constraints and is maximally consistent with the channel observations.
\end{theorem}

\section{Experimental Results}
In this section, we present the experimental results of the decoding scheme.
To clearly demonstrate the performance of the proposed approach, we first describe the experimental setup used in our simulations. This is followed by a detailed discussion of the experimental results.

\renewcommand{\figurename}{FIGURE}
\begin{figure*}[!t]
\centering
\begin{minipage}{0.65\columnwidth}
\centering
\includegraphics[width=\columnwidth]{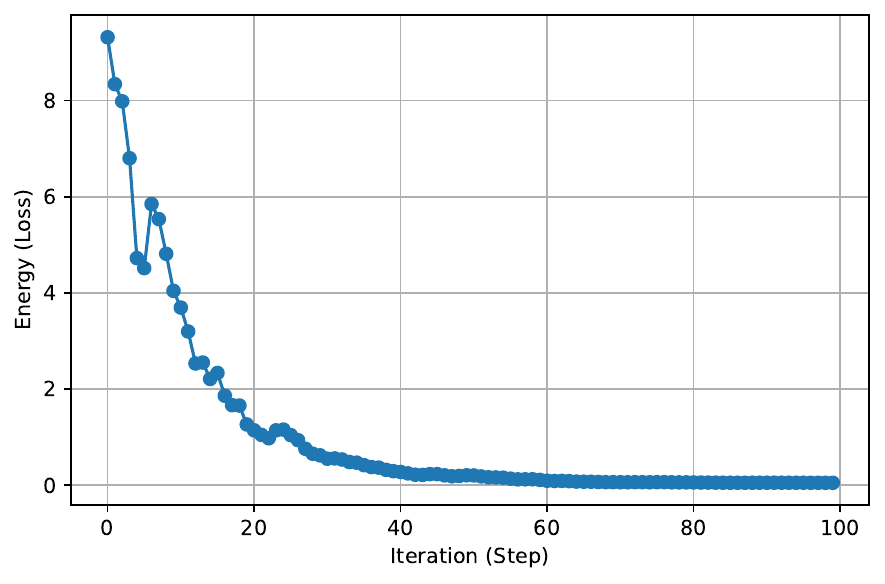}
\end{minipage}
\begin{minipage}{0.65\columnwidth}
\centering
\includegraphics[width=\columnwidth]{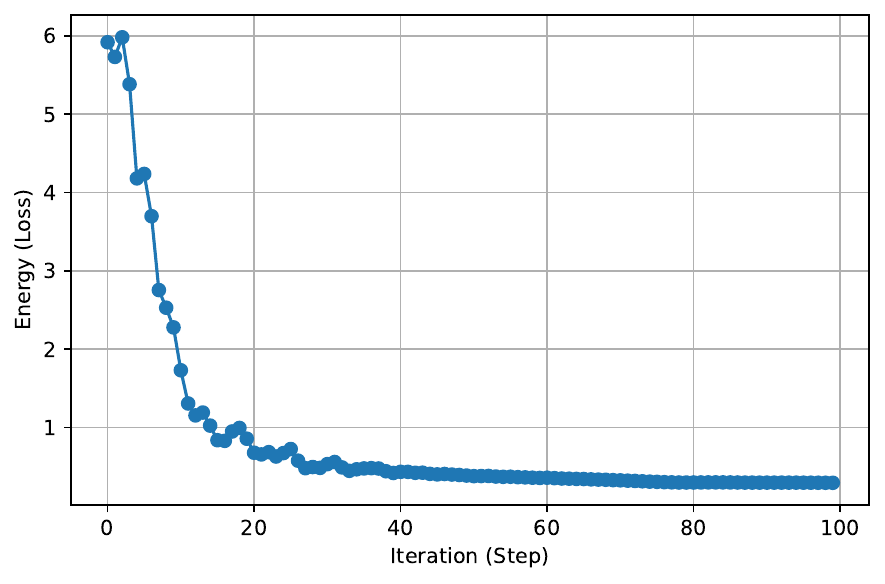}
\end{minipage}
\begin{minipage}{0.65\columnwidth}
\centering
\includegraphics[width=\columnwidth]{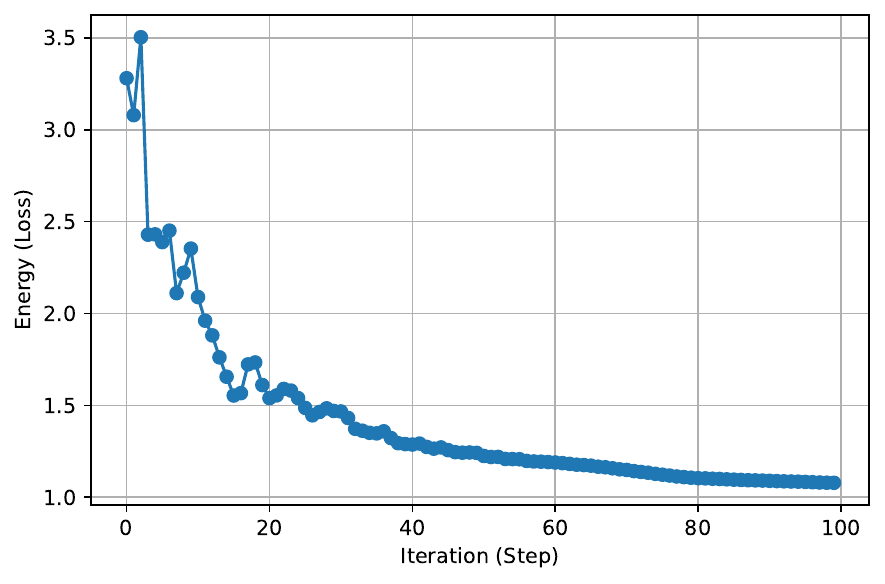}
\end{minipage}
\caption{Energy convergence of the QAOA-based decoder versus optimization steps for three different experimental settings ($\sigma = 1, 1.5, 2$).}
\label{fig:qaoa_energy_side}
\end{figure*}

\subsection{Experimental Configuration}

All experiments are conducted under the assumption of Binary Phase Shift Keying (BPSK)~\cite{proakis2001digital} transmission over an Additive White Gaussian Noise (AWGN) channel~\cite{proakis2001digital}.
Let the transmitted binary codeword for \( n \) code length given by
\begin{align*}
    \mathbf{c} := [c_0, c_1, \ldots, c_{n-1}], \quad c_i \in \{0,1\}.
\end{align*}
The binary symbols are modulated using BPSK, where each bit is mapped according to $x_i = (-1)^{c_i},$
corresponding to the mapping \( 0 \mapsto +1 \) and \( 1 \mapsto -1 \). The resulting transmitted signal vector is therefore given by
\begin{align*}
    \mathbf{x} = [(-1)^{c_0}, (-1)^{c_1}, \ldots, (-1)^{c_{n-1}}].
\end{align*}
The modulated signal is transmitted over an AWGN channel, where the received signal is expressed as 
\begin{align*}
    y_i = x_i + n_i.
\end{align*}
The noise samples \(n_i\) are assumed to be independent and identically distributed Gaussian random variables with zero mean and variance \(\sigma^2\), i.e., $n_i \sim \mathcal{N}(0, \sigma^2).$
The corresponding Gaussian probability density function~\cite{richardson2008modern} is given by
\begin{align*}
    p(n) = \frac{1}{\sqrt{2\pi\sigma^2}}
\exp\left(-\frac{n^2}{2\sigma^2}\right).
\end{align*}
For BPSK transmission over an AWGN channel, soft information is provided to the decoder in the form of LLRs. The LLR associated with the received symbol \(y_i\) is given by $ \mathrm{LLR}_i  = \tfrac{2}{\sigma^2} y_i.$
These LLR values are used as inputs to both the classical belief propagation decoder and the proposed QAOA-based decoding framework.

\subsection{Results and Discussion}
In this section, we demonstrate through numerical simulations that the proposed QAOA-based decoder achieves improved performance compared to the BP algorithm.
For simulating quantum circuits, we use PennyLane~\cite{bergholm2018pennylane}, while the framework of the architecture is constructed using PyTorch~\cite{paszke2019pytorch}.
In the first row of Table~\ref{tab:bp_qaoa_results}, we report the decoding performance for an LDPC code with parameters \( [n, k] = [6, 2] \). The corresponding parity-check matrix is given by 
\begin{equation*}
H =
\begin{pmatrix}
1 & 1 & 1 & 0 & 0 & 0 \\
1 & 1 & 0 & 1 & 0 & 0 \\
0 & 0 & 1 & 1 & 1 & 0 \\
0 & 0 & 1 & 1 & 0 & 1
\end{pmatrix}.
\end{equation*}
The null space of \(H\) has dimension \(k = 2\), and a basis for this null space is chosen as $\mathbf{v}_1 = [1, 0, 1, 1, 0, 0], \quad
\mathbf{v}_2 = [1, 1, 0, 0, 0, 0].$
Using this basis, the complete set of valid codewords is generated.
For classical BP decoding, the maximum number of iterations is fixed to 100. In the quantum-assisted decoding framework, the QAOA is employed with circuit depth \(l = 10\).
\begin{table}
\centering
\caption{Performance comparison of BP and QAOA-based decoder under different noise levels.}
\label{tab:bp_qaoa_results}
\begin{tabular}{cccccc}
\hline
$[n,k]$ & $|C|$ & $\sigma$ & $|Y|$ & BP~\cite{richardson2008modern} & QAOA~(This Work) \\
\hline
\multirow{3}{*}{$[6,2]$} & \multirow{3}{*}{4}
 & 1.0 & \multirow{3}{*}{25} & 0.82 & \textbf{0.85} \\
 & & 1.5 & & 0.56 & \textbf{0.66} \\
 & & 2.0 & & 0.47 & \textbf{0.56} \\
\hline
\multirow{3}{*}{$[7,3]$} & \multirow{3}{*}{8}
 & 1.0 & \multirow{3}{*}{13} & 0.76 & \textbf{0.82} \\
 & & 1.5 & & 0.50 & \textbf{0.64} \\
 & & 2.0 & & 0.31 & \textbf{0.61} \\
\hline
\multirow{3}{*}{$[8,4]$} & \multirow{3}{*}{16}
 & 1.0 & \multirow{3}{*}{6} & 0.67 & \textbf{0.81} \\
 & & 1.5 & & 0.37 & \textbf{0.61} \\
 & & 2.0 & & 0.32 & \textbf{0.58} \\
\hline
\end{tabular}
\end{table}

To evaluate performance under different noise conditions, the noise standard deviation is fixed at \(\sigma \in \{1, 1.5, 2\}\). For each value of \(\sigma\), a total of 100 independent trials are conducted. Let \(|C|\) denote the total number of valid codewords, and let \(|Y|\) denote the total number of noisy outputs generated per codeword after transmission through the noise channel, corresponding to the number of independent noise realizations used for testing. The last two columns of Table~\ref{tab:bp_qaoa_results} report the empirical probability of successful decoding, defined as the ratio of correctly decoded outputs to the total number of tested noisy instances. We observe that as the noise standard deviation $\sigma$ increases, the probability of successful decoding using the BP algorithm decreases. In contrast, the QAOA-based decoder maintains a higher decoding success probability than the BP decoder.

To further demonstrate the generality of the proposed QAOA-assisted LDPC decoding framework, we consider two additional linear block codes with parameters \([n,k] = [7,3]\) and \([n,k] = [8,4]\). 
For the \([7,3]\) and \([8,4]\) codes, the total numbers of independent trials are set to 104 and 96, respectively. As in the previous experiments, the noise standard deviation is fixed to \(\sigma \in \{1, 1.5, 2\}\), while all other experimental parameters are kept unchanged to ensure consistent performance comparison across different code lengths.


Figure~\ref{fig:qaoa_energy_side} shows the energy convergence of the QAOA-based decoder as a function of the classical optimization iterations for three different noise levels. The experiments are conducted for a linear block code with parameters \([n, k ]= [6, 2]\), whose parity-check matrix is given by 
\begin{equation}\nonumber
H =
\begin{pmatrix}
1 & 1 & 1 & 0 & 0 & 0 \\
0 & 1 & 1 & 1 & 0 & 0 \\
0 & 0 & 1 & 1 & 1 & 0 \\
0 & 0 & 0 & 1 & 1 & 1
\end{pmatrix}.
\end{equation}
The QAOA circuit depth is fixed to \(l = 10\).
The three subplots correspond to noise standard deviations \(\sigma = 1\), \(1.5\), and \(2\), respectively, while the total number of classical optimization steps is set to 100. As observed, the QAOA energy decreases consistently with the number of iterations, indicating stable convergence of the variational parameters. 

\section{Advantages}
BP and QAOA are two completely different decoding concepts.
BP is a local iterative message-passing algorithm that greedily updates beliefs based on neighboring parity-check constraints and channel knowledge.
This leads to BP being based on local consistency and potentially converging to suboptimal fixed points, especially in the case of short cycles or trapping sets.
QAOA is an optimization-based decoding method that formulates the decoding problem as an optimization task by expressing both parity-check constraints and channel reliability knowledge in a unified cost Hamiltonian.
QAOA leverages quantum superposition and parameterized unitary evolution to explore a vast configuration space of candidate bit strings simultaneously, instead of pursuing local updates.
Therefore, QAOA offers a complementary decoding solution that is especially useful in decoding instances where classical iterative decoders demonstrate poor convergence properties.

An important aspect of the proposed decoding framework is its computational complexity. For comparison, the BP algorithm admits linear-time complexity with respect to the code length under standard assumptions. In contrast, the QAOA-based decoder involves both quantum state evolution and classical parameter optimization.
On the classical side, a momentum-based gradient descent method is employed to optimize the variational parameters \( \boldsymbol{\gamma} \) and \( \boldsymbol{\beta} \) appearing in the parameterized unitary operations. The computational complexity of this classical optimization procedure is \( O(\mathrm{poly}(m)) \), where \( m \) denotes the number of optimization iterations.
On the quantum side, the dominant computational cost arises from the repeated preparation and evolution of the quantum state according to the QAOA circuit. As discussed in Ref.~\cite{zhou2020quantum, zhang2022applying}, the corresponding complexity \( O(\mathrm{poly}(l)) \), where \( l \) is the number of layers.
Consequently, the overall computational complexity of the proposed QAOA-based decoding scheme can be expressed as $ O\bigl(\mathrm{poly}(l) + \mathrm{poly}(m)\bigr)$.
The QAOA procedure runs in polynomial time with respect to the code length and the circuit depth, respectively. However, the QAOA-based decoder is observed to provide better decoding performance than the BP algorithm.

\section{Conclusion and Future Directions}

In this study, we explored a QAOA-based decoding framework for LDPC codes by formulating the classical decoding objective as a quantum cost Hamiltonian.
The proposed approach casts LDPC decoding as an optimization problem, in which parity-check constraints and channel reliability information are jointly encoded and optimized using the QAOA.
Our experimental results, obtained on different problem instances, demonstrate that the QAOA-based decoder achieves performance comparable to BP in standard decoding scenarios.
The observed results indicate that optimization-based quantum decoding can provide meaningful advantages even in the near-term regime.

In future work, we would like to extend the proposed framework by exploring more expressive QAOA tailored to the structural properties of LDPC decoding, to improve decoding performance, while reducing computational complexity overhead.  
In particular, we develop theoretical performance and investigate their behavior in the large block-length regime. 
These research directions are essential for assessing the practical potential and scalability of quantum optimization methods for error-correcting code decoding.



{
\bibliographystyle{IEEEtran}
\bibliography{IEEEabrv,references}
}
\end{document}